\documentclass[12pt]{article}
\usepackage{color,graphicx} %showkeys, 
\usepackage[mathcal]{euscript}
\usepackage{graphicx}
\usepackage{amssymb}
\usepackage{epstopdf}
\usepackage{latexsym}
\usepackage{indentfirst}
\def\Tr{{\rm Tr\,}}

\def\K3{\mathrm K3}

\def\double #1{#1{\hbox{\kern-2pt $#1$}}}
\def\p {\partial}
\def\half {{1\over 2}}

\def\a{{\alpha}}
\def\ah{{\hat\alpha}}

\def\d{{\delta}}

\def\ad{{\dot{\alpha}}}
\def\bd{{\dot{\beta}}}

\newcommand{\beq}{\begin{equation}}
\newcommand{\eeq}{\end{equation}}
\newcommand{\bea}{\begin{eqnarray}}
\newcommand{\eea}{\end{eqnarray}}
\def\p{{\partial}}
\def\pb{{\bar\partial}}
\def\ah{{\hat\alpha}}

\def\N{{\cal N}}

\def\Tb{{\mathbf{T}}}

\newcommand{\w}{\widehat}

\def\J{{\mathsf J}}

\def\a{{\alpha}}
\def\ah{{\widehat\alpha}}
\def\ad{{\dot a}}
\def\bd{{\dot b}}
\def\l{{\lambda}}
\def\lh{{\widehat\lambda}}

\def\d{{\delta}}

\def\ve{{\varepsilon}}
\def\veb{\overline\varepsilon}
\def\s{{\sigma}}

\def\s{{\sigma}}

\def\N{{\nabla}}
\def\Nb{{\overline\nabla}}
\def\Nh{{\widehat N}}

\def\o{{\omega}}
\def\half{{1\over 2}}
\def\p{{\partial}}
\def\pb{{\overline\partial}}

\def\oh{{\widehat\o}}

\def\Qh{{\widehat Q}}
\def\add{{\dot a}}
\def\bdd{{\dot b}}
\def\cdd{{\dot c}}

\def\Jb{\overline J}
\def\jb{\overline j}

\def\Tb{\overline T}

\def\pp{{\mathchoice
          {%displaystyle
              \kern 1pt%
              \raise 1pt
              \vbox{\hrule width5pt height0.4pt depth0pt
                    \kern -2pt
                    \hbox{\kern 2.3pt
                          \vrule width0.4pt height6pt depth0pt
                          }
                    \kern -2pt
                    \hrule width5pt height0.4pt depth0pt}%
                    \kern 1pt
           }
            {%textstyle
              \kern 1pt%
              \raise 1pt
              \vbox{\hrule width4.3pt height0.4pt depth0pt
                    \kern -1.8pt
                    \hbox{\kern 1.95pt
                          \vrule width0.4pt height5.4pt depth0pt
                          }
                    \kern -1.8pt
                    \hrule width4.3pt height0.4pt depth0pt}%
                    \kern 1pt
            }
            {%scriptstyle
              \kern 0.5pt%
              \raise 1pt
              \vbox{\hrule width4.0pt height0.3pt depth0pt
                    \kern -1.9pt  %[e]=0.15pt
                    \hbox{\kern 1.85pt
                          \vrule width0.3pt height5.7pt depth0pt
                          }
                    \kern -1.9pt
                    \hrule width4.0pt height0.3pt depth0pt}%
                    \kern 0.5pt
            }
            {%scriptscriptstyle
              \kern 0.5pt%
              \raise 1pt
              \vbox{\hrule width3.6pt height0.3pt depth0pt
                    \kern -1.5pt
                    \hbox{\kern 1.65pt
                          \vrule width0.3pt height4.5pt depth0pt
                          }
                    \kern -1.5pt
                    \hrule width3.6pt height0.3pt depth0pt}%
                    \kern 0.5pt%}
            }
        }}

\def\mm{{\mathchoice
                       {%displaystyle
                             \kern 1pt
               \raise 1pt    \vbox{\hrule width5pt height0.4pt depth0pt
                                  \kern 2pt
                                  \hrule width5pt height0.4pt depth0pt}
                             \kern 1pt}
                       {%textstyle
                            \kern 1pt
               \raise 1pt \vbox{\hrule width4.3pt height0.4pt depth0pt
                                  \kern 1.8pt
                                  \hrule width4.3pt height0.4pt depth0pt}
                             \kern 1pt}
                       {%scriptstyle
                            \kern 0.5pt
               \raise 1pt
                            \vbox{\hrule width4.0pt height0.3pt depth0pt
                                  \kern 1.9pt
                                  \hrule width4.0pt height0.3pt depth0pt}
                            \kern 1pt}
                       {%scriptscriptstyle
                           \kern 0.5pt
             \raise 1pt  \vbox{\hrule width3.6pt height0.3pt depth0pt
                                  \kern 1.5pt
                                  \hrule width3.6pt height0.3pt depth0pt}
                           \kern 0.5pt}
                       }}

\def\ad{{\kern0.5pt
                   \alpha \kern-5.05pt
\raise5.8pt\hbox{$\textstyle.$}\kern 0.5pt}}

\def\bd{{\kern0.5pt
                   \beta \kern-5.05pt \raise5.8pt\hbox{$\textstyle.$}\kern 0.5pt}}

\def\qd{{\kern0.5pt
                   q \kern-5.05pt \raise5.8pt\hbox{$\textstyle.$}\kern 0.5pt}}
\def\Dot#1{{\kern0.5pt
     {#1} \kern-5.05pt \raise5.8pt\hbox{$\textstyle.$}\kern 0.5pt}}
\begin{document}

\setcounter{page}0
\thispagestyle{empty}
\begin{flushright}
\makebox[0pt][b]{}
\end{flushright}
\vspace{30pt}

\begin{center}
{\LARGE $AdS$ pure spinor superstring in constant backgrounds}\\

\vspace{30pt}
{Osvaldo Chandia${}^{\diamondsuit}$\footnote{email: ochandiaq@gmail.com}, L. Ibiapina Bevilaqua${}^{\clubsuit}$\footnote{email: leandro@ect.ufrn.br}and Brenno Carlini
Vallilo${}^{\spadesuit}$\footnote{email: vallilo@unab.cl}
}
\vspace{30pt}

${}^{\diamondsuit}${\em Departamento de Ciencias, Facultad de Artes Liberales, Universidad Adolfo Ib\'a\~nez,\\ Facultad de Ingenier\'{\i}a y Ciencias, Universidad Adolfo Ib\'a\~nez,\\ Diagonal Las Torres 2640, Pe\~nalol\'en, Santiago, Chile}

${}^{ \clubsuit}${\em Escola de Ci\^{e}ncias e Tecnologia, Universidade Federal do Rio Grande do Norte,\\ Caixa Postal 1524, 59072-970, Natal, RN, Brazil}

${}^{ \spadesuit}${\em Facultad de Ciencias Exactas, Departamento de Ciencias F\'{\i}sicas, \\ 
Universidad Andres Bello, Republica 220, Santiago, Chile}
\vspace{60pt}

 {\bf Abstract}
\end{center}
In this paper we study the pure spinor formulation of the superstring in $AdS_5\times S^5$ around point particle 
 solutions of the classical equations of motion. As a particular example we quantize the pure spinor string 
 in the BMN background. 

\newpage

{ \parskip=0.0in \tableofcontents }

%%%%%%%%%%%%%%%%%%

\section{Introduction}

Although a lot of work has been done on strings on $AdS_5\times S^5$, no
complete quantization of the model has been performed. The classical
integrability found in the Green-Schwarz \cite{bena} and pure spinor
\cite{vallilo} formalisms, has not been extended to the full quantum theory.
Interesting work in this direction was done by Benichou
\cite{benichou}, where the pure spinor formalism was used to derive the Y-system
equations \cite{gkv} using world sheet techniques.

One particular advantage of the pure spinor formalism is that it can
be used without fixing any gauge, so conformal field theory methods 
(and conformal perturbation theory) can be used \cite{Berkovits:2000fe,Berkovits:2000yr}. Quantum conformal
invariance for any on shell classical background was proven in 
\cite{oneloop, quanconsistency}. Other quantum consistency checks for
the $AdS$ pure spinor string were done in
\cite{radius,algebra}. Furthermore, backgrounds that do not admit
light-cone gauge fixing, the pure spinor description can be used. The formalism
was recently  used \cite{konishi}\footnote{For related work, see
  \cite{radu, gromov}.} to compute the energy of a particular 
string state conjectured to be dual to one element of the Konishi
multiplet. The result agreed with the strong coupling computation
using the Y-system \cite{gkv2}. Some aspects of semiclassical quantization of the $AdS$ pure spinor superstring have been discussed in \cite{Aisaka:2012ud,Cagnazzo:2012uq,Tonin:2013uec}.

In this paper we use the background field method to study the pure
spinor superstring in constant backgrounds.  The method is the same as 
the one used in \cite{konishi}, although more details will be presented. 
The particular case of a BPS background \cite{bmn} is given as an example. 
Unlike in \cite{konishi}, the BMN limit can be used and hence the full spectrum of the string can be 
computed. As expected, it agrees with light-cone GS formalism. The
approach taken here is very different from a previous description of
the pp-wave background using pure spinors \cite{ppnathan}. In that
work, the background is described by a contraction of the
${\mathfrak{psu}}(2,2|4)$ algebra and hence is nonlinear in the
world-sheet variables. We will see that the background field method
gives a much simpler description. We will also see that the part of
the spectrum related to the supergravity multiplet has a very simple 
description, almost exactly as in the light-cone GS quantization
\cite{metsaev}.  This is a surprise because massless vertex operators 
for the pure spinor string in $AdS$ are not known in general. For
progress in this direction, see
\cite{mikhailov1,mikhailov2,mikhailov3,Chandia:2013kja}. It would be interesting to
study the spectrum we obtain away from the BMN limit. 

In section 2 we review the construction of the string on a $AdS_5\times S^5$ background using the pure spinor formalism. In section 3 we expand the world-sheet action around a bosonic solution of the classical equations of motion and verify how the BRST and conformal transformations are compensated because of a gauge fixing and such that the action remains invariant. Finally, in section 4 we study the special case of the BPS background, in particular, we study the spectrum in this case.

\section{$AdS$ Pure Spinor String}

In this section we present a short review of the pure spinor formalism
for the $AdS$ background and describe the BPS background used in the
rest of the paper.

It is well-known that the $AdS_5\times S^5$ background is described by
the coset $\frac{PSU(2,2|4)}{SO(1,4)\times SO(5)}$. Elements of this
coset are described by $g\in PSU(2,2|4)$ and two elements are
identified if they differ by local right multiplication by en element
of $SO(1,4)\times SO(5)$. Classical solutions in this background are
elements of the coset $g(\tau,\sigma)$ satisfying the equations of
motion that come from the action plus the classical Virasoro
condition. In this section we use the notation $z=\tau-\sigma$, $\bar
z=\tau +\sigma$ and $\sigma$ is periodic with period $2\pi$, and we use the world-sheet derivatives $\p=\frac12 (\p_\tau-\p_\s), \pb=\frac12 (\p_\tau+\p_\s)$.

The action in the pure spinor formalism is given by \cite{Berkovits:2000yr,quanconsistency}
\beq\label{psaction} S= \left< \half J_2 \Jb_2 + {3\over 4}
J_3 \Jb_1  + {1\over 4}
J_1 \Jb_3 +
\omega  \Nb \lambda  +
\overline\omega \nabla \overline\lambda  - N \overline N \right> ,
\eeq where
\beq 
\left< \cdots \right> = \frac{1}{\pi\a'} \int d^2z \Tr ,
\eeq
and, thanks to the $\mathbb{Z}_4$ grading of the $\mathfrak{psu}(2,2|4)$ Lie algebra\footnote{The Lie algebra ${\cal H}$ is decomposed as  ${\cal H}_0+ {\cal H}_1+ {\cal H}_2+ {\cal H}_3$. Note that the commutators in the Lie algebra map $ {\cal H}_i\times {\cal H}_j$ into $ {\cal H}_{i+j}$ mod $4$. }, the Metsaev-Tseytlin currents $J = g^{-1}\p g$ have components  
\beq J_0 = (g^{-1}\p g)^{[mn]} T_{[mn]},\quad
J_1 = (g^{-1}\p g)^{\alpha} T_{\alpha},\quad
J_2 = (g^{-1}\p g)^{m} T_{m},\quad
J_3 = (g^{-1}\p g)^{\ah} T_{\ah},
\eeq
where $\{ T_{[mn]}, T_m, T_\alpha, T_{\hat\alpha} \}$ are the algebra generators. Later we will use a more convenient basis. The non zero commutators can be found in the Appendix. Similarly we define $\Jb = g^{-1} \pb g$. The pure spinor variables are defined as 
\beq
\omega = \omega_\a T_\ah \d^{\a\ah},
\quad\lambda = \lambda^\a T_\a,\quad N =  -\{\omega,\lambda\}, 
\eeq
\beq
\overline\omega = \overline\omega_\ah T_\a \d^{\a\ah},
\quad\overline\lambda = \overline\lambda^\ah T_\ah,\quad \overline N =
-\{\bar\omega,\bar\lambda\}.
\eeq
The world-sheet covariant derivatives in the action are defined as 
$$\nabla  = \partial  + [J_0, \;\; ], \quad
\overline\nabla  = \overline\partial  + [\Jb_0,\;\; ] .$$

The BRST charge is given by
\beq Q = \oint d\sigma \Tr [\lambda J_3 +\overline\lambda \bar J_1] .\eeq
% The equations of motion we get from this action are\footnote{There are other ways to write these equations using the Muerer-Cartan equations $\partial \bar J - \bar\partial J +[J,\bar J]=0$.}
% \beq   \bar\nabla J_2 +\nabla \bar J_2 +[J_1,\bar J_1]- [J_3,\bar J_3]+ [N,\bar J_2] +[\bar N,J_2]=0
% \eeq
% $$ \frac{1}{2}\nabla \bar J_1 + \frac{3}{2}\bar\nabla J_1+...  =0
% $$
% $$\frac{3}{2}\nabla \bar J_3 + \frac{1}{2}\bar\nabla J_3+... =0
% $$
% $$\bar\nabla \lambda -[ \bar N ,\lambda]=0,\quad \nabla \bar\lambda -[ N ,\bar \lambda]=0$$
%The current corresponding to the global symmetry $\delta g = \Omega g$, where $\Omega$ is a constant element of $PSU(2,2|4)$ is calculated by the Noether method and is given by
%\beq
%j = \frac{1}{2\pi\alpha'} g[  J_2  + \frac{1}{2} J_1 + \frac{3}{2} J_3  + N ]g^{-1}
%\eeq
%$$\bar j = \frac{1}{2\pi\alpha'}  g [\bar J_2  + \frac{3}{2} \bar J_1 + \frac{1}{2}\bar J_3 +\bar N  ]g^{-1}
%$$
%This current is conserved $\bar\partial j +\partial \bar j=0$ using the equations of motion. 
%Note that this current is {\em not} $(\partial g) g^{-1}$. The time component, the one that enters in the charge, is
%\beq
%j_\tau = \frac{1}{2\pi\alpha'} g[  (J_2)_\tau  +  (J_1)_\tau +  (J_3)_\tau  + (J_1)_\sigma - (J_3)_\sigma + N +\bar N]g^{-1}.
%\eeq 

We study classical solutions with vanishing BRST charge. A simple way
to get this is with $J_3=\overline J_1=0$ or with vanishing $\l$ and $\overline\l$. 

The theory is also constrained by the conformal symmetry which is generated by the stress tensor. The classical stress tensor is given by
\beq\label{stress} T = \frac{1}{\alpha'}\Tr [J_2 J_2 + 2J_1J_3 + \omega \N \lambda],\quad \overline T =\frac{1}{\alpha'}\Tr [\Jb_2 \Jb_2 + 2\Jb_1\Jb_3 + \overline\omega \Nb\overline\lambda].\eeq

Using the world-sheet equations of motion, it can be shown that $T$ is holomorphic and $\Tb$ is antiholomorphic. The conformal transformation of the group element $g$ is
\beq\label{Tg} \d g = \ve \p g + \veb \pb g ,
\eeq
where $\ve$ is a holomorphic parameter and $\veb$ is an antiholomorphic parameter. The pure spinor variables transform as
\beq\label{Tonpure} \d\l = \ve\p\l + \veb\pb\l,\quad \d\o = \p(\ve\o) + \veb\pb\o\eeq
$$
\d\lh = \ve\p\lh + \veb\pb\lh,\quad \d\oh = \pb(\veb\oh) + \ve\p\oh .$$
It is possible to show that, under (\ref{Tg}) and (\ref{Tonpure}), the action (\ref{psaction}) is invariant. A similar calculation shows that, using Noether theorem, (\ref{stress}) are the conserved quantities determined by this symmetry.

\section{Background Field Expansion}

We are interested in $\sigma$ independent classical solutions which are not necessarily BPS. For instance, the action (\ref{psaction}) has a classical non BPS bosonic solution with energy ${\cal E}$ and angular momentum ${\cal J}$  in one direction of $S^5$. The solution is given by the group element 
\beq\label{sol} g=e^{\alpha'\tau(-{\cal E} T + {\cal J} J) } ,\eeq
where $\tau$ is the time coordinate of the world-sheet. Here $T$ is the translation along the time-like direction of $AdS_5$ and $J$ is a translation along a great equator of $S^5$. Note that both $T$ and $J$ belong to ${\cal{H}}_2$. The pure spinor variables are zero for this solution.  

To show that ${\cal E}$ and ${\cal J}$ are the energy and the angular momentum of the classical solution, we first find the conserved current due to the global $PSU(2,2|4)$ symmetry of the action (\ref{psaction}). Infinitesimally, we have $\d g= \ve g$ and we follow the Noether procedure to find the conserved current. That is, we assume $\ve$ to be non-constant such that the action varies with a term linear in the derivative of $\ve$. Varying the action (\ref{psaction}) we obtain
\beq\label{dS} \d S = \langle \p\ve \jb + \pb\ve j \rangle ,\eeq
where
\beq\label{jeis} j = g ( \half J_2 + {3\over 4} J_3 + {1\over 4} J_1 + N ) g^{-1},\quad \jb = g ( \half \Jb_1 + {1\over 4} \Jb_3 + {3\over 4} \Jb_1 + \Nh ) g^{-1} .\eeq
For the solution (\ref{sol}) the non-zero currents are $J_2=\Jb_2 = (-{\cal E} T + {\cal J} J)$. We have that the energy and the angular momentum of the classical string configuration are given by
\beq\label{detab} {\cal E}={1\over 2\pi\alpha'}\oint d\s {\rm Str}(Tj_\tau) ,\eeq
$$
{\cal J}={1\over 2\pi\alpha'}\oint d\s {\rm Str}(J j_\tau).$$
Note that the classical Virasoro constraint is violated if ${\cal E} \neq |{\cal J}|$ 
$$T=\bar T = {\alpha'\over 4}(-{\cal E}^2+ {\cal J}^2) .$$ 

We now expand around the classical solution  $g=g_0 e^{\sqrt{\alpha'}X}$, where $X$ belongs to ${\cal H}_1 + {\cal H}_2 + {\cal H}_3$. This is a gauge choice. This gauge choice does not preserve  the other classical symmetries. Consider a BRST transformation 
$$
Q_B g = g ( \l + \lh ) .$$
From here, we obtain
\beq\label{QX} e^{-\sqrt{\alpha'}X} Q_B e^{\sqrt{\alpha'}X} = \l + \lh ,\eeq
and expanding the left hand side,
\beq\label{QXexp} \sqrt{\alpha'}Q_B X +\alpha' \half [ Q_B X , X ] + \cdots  = \l + \lh.\eeq
We try to determine the BRST transformation of $X$ as an expansion in powers of $X$, that is
\beq\label{QXpert} Q_B X ={1\over \sqrt{\alpha'}} \sum_{n=0} (Q_B)_n X ,\eeq
where $(Q_B)_n X$ contains $n$ powers of $\sqrt{\alpha'}X$. Plugging this expansion in (\ref{QX}) we obtain that
\beq\label{QXX} Q_B X = {1\over\sqrt{\alpha'} }( \l + \lh ) - \half [ ( \l + \lh ) , X ] + \cdots. \eeq
Since the original $X$ does not have an ${\cal H}_0$ component, and $[ \l , X_3 ] + [ \lh , X_1 ]$ belongs to ${\cal H}_0$,  this BRST transformation does not preserve the form of $g$. Note that this analysis does not depend on the form of the background. To preserve 
our initial gauge choice we have to perform a compensating $SO(1,4)\times SO(5)$ gauge transformation. The $SO(1,4)\times SO(5)$ gauge transformations for $X_0$ is given by $\delta X_0 ={1\over\sqrt{\alpha'}} \Lambda $, where $\Lambda \in SO(1,4)\times SO(5)$. The ${\cal H}_0$ component in 
(\ref{QXX}) can be eliminated if we choose $\Lambda= {\sqrt{\alpha'}\over 2} [  \l  , X_3 ] + {\sqrt{\alpha'}\over 2} [ \lh  , X_1 ]+\cdots$. This compensating gauge transformation will affect all fields, since now the BRST transformation should come together with it. In particular, the pure spinor ghosts now have a non vanishing BRST transformation. 

A similar problem happens when we consider the conformal transformation of the gauge group element (\ref{Tg}) which implies
\beq\label{eXdeX} e^{-\sqrt{\alpha'}X} \d e^{\sqrt{\alpha'}X} = \ve ( e^{-\sqrt{\alpha'}X} \p e^{\sqrt{\alpha'}X} + e^{-\sqrt{\alpha'}X} J_{(0)} e^{\sqrt{\alpha'}X} ) + \veb (  e^{-\sqrt{\alpha'}X} \pb e^{\sqrt{\alpha'}X} + e^{-\sqrt{\alpha'}X} \Jb_{(0)} e^{\sqrt{\alpha'}X} ) ,\eeq
where $J_{(0)}$ and $\Jb_{(0)}$ are the components of the Metsaev-Tseytlin currents evaluated in the background (\ref{sol}). Note that these currents have values in ${\cal H}_2$ only. Expanding the exponentials in (\ref{eXdeX}) we have
\beq\label{exx} \sqrt{\alpha'}\d X +\alpha' \half [\sqrt{\a'} d X , X ] + \cdots = \ve ( J_{(0)} + \sqrt{\alpha'}\p X + \sqrt{\alpha'}[ J_{(0)} , X ] + \cdots ) +  \veb ( J_{(0)} + \sqrt{\alpha'}\pb X + \sqrt{\alpha'}[ J_{(0)} , X ] + \cdots ) .\eeq
As before, we try to solve this equation in powers of $X$ such that 
\beq\label{expdX} \d X ={1\over \sqrt{\alpha'}} \sum_{n=0} \d_n X ,\eeq
where $\d_n X$ contains $n$ powers of $\sqrt{\alpha'}X$. Then, up to first order in $X$,
\beq\label{dX} \d X = \ve ( {1\over \sqrt{\alpha'}}J_{(0)} + \p X + \half [ J_{(0)} , X ] + \cdots ) + \veb ({1\over\sqrt{\alpha'}} J_{(0)} + \pb X + \half [ J_{(0)} , X ] + \cdots ) .\eeq
Since the original $X$ does not have an ${\cal H}_0$ component, and $[ J_{(0)} , X_2 ]$ belongs to ${\cal H}_0$,  this conformal transformation does not preserve the gauge choice for $X$. The compensating gauge transformation to restore the gauge choice for $X$ in this case is $\Lambda=-\sqrt{\alpha'}{\ve\over 2}[ J_{(0)} , X_2 ]-\sqrt{\alpha'}{\bar\ve\over 2}[ J_{(0)} , X_2 ]+ \cdots $, where $\cdots$ are higher order terms in $\alpha'$. Another way to obtain these transformations is to use canonical quantization. In this case, the momenta conjugate to the fields have higher order corrections and these corrections are used to replace time derivatives of $X$.

\subsection{ Expansion of the action}

Before expanding the action, we need to determine up to what order we will expand it to check BRST and conformal invariance. Generically, the action is expanded as
\beq\label{expS} S = S_0 + S_1 + S_2 + S_3 + \cdots,\eeq
where $S_n$ is of order $n$ in $(X, \l, \o, \lh, \oh)$. Note that $S_0$ is the value of the action for the background and $S_1$ vanishes because the background satisfies the classical equations of motion.

Consider the BRST symmetry first. The transformation for the world-sheet fields are given by (\ref{QXX}). Note that they start with a term linear in $(X, \l, \o, \lh, \oh)$, then $Q_B = (Q_B)_0 + (Q_B)_1 + \cdots$. When we act the BRST generator on the action, we can expand $Q_B S$ as an expansion in powers of $(X, \l, \o, \lh, \oh)$ as
\beq\label{QS}Q_B S =  (Q_B)_0 S_2 + \left( (Q_B)_0 S_3 + (Q_B)_1 S_2 \right) + \cdots .\eeq
The invariance of the action order by order implies
\beq\label{QSord}  (Q_B)_0 S_2 = 0,\quad  \left( (Q_B)_0 S_3 + (Q_B)_1 S_2 \right)  = 0.\eeq
Then, if we want to test BRST invariance up to the the order shown above, then the action has to be expanded up to order three in $(X, \l, \o, \lh, \oh)$. 

Consider now the conformal symmetry.  The transformation for the world-sheet fields are given by (\ref{Tonpure}) and (\ref{dX}). Note that they start with a term independent of $X$ in (\ref{dX}), then $\d = \d_0 + \d_1 + \cdots$. When we vary the action under the conformal transformations, $\d S$ can be written as an expansion in powers of $(X, \l, \o, \lh, \oh)$ as
\beq\label{deltaS} \d S =  \d_0 S_2 + \left( \d_1 S_2 + \d_0 S_3 \right) + \cdots .\eeq
The action is invariant order by order implies
\beq\label{dSord}   \d_0 S_2 = 0,\quad  \left( \d_1 S_2 + \d_0 S_3\right)  = 0.\eeq
Then, if we want to test conformal invariance up to the the order shown above, we are again led to the conclusion that the action the action has to be expanded up to order three in $(X, \l, \o, \lh, \oh)$.

We now expand the action up to order three in the quantum fields $(X, \l, \o, \lh, \oh)$. We expand the action (\ref{psaction}) around the background given in (\ref{sol}), that is $g=g_0 e^{\sqrt{\alpha'}X}$ with $g_0$ given as in (\ref{sol}) and $X =  X_1 + X_2 + X_3$. This solution is such that $J_{(0)}=\Jb_{(0)}$.  The Metsaev-Tseytlin current becomes
\beq\label{expJ} J=g^{-1} dg = e^{-\sqrt{\alpha'}X}J_{(0)} e^{\sqrt{\alpha'}X}+e^{-\sqrt{\alpha'}X}d e^{\sqrt{\alpha'}X} ,\eeq
where $d$ represents $\p$ and $\pb$. Expanding, we obtain
\beq\label{JJJ} J = J_{(0)} + \sqrt{\alpha'}J_{(1)} +{\alpha'\over 2} J_{(2)}+{{\alpha'}^{3\over 2}\over 6} J_{(3)} + \cdots ,\eeq
where
$$
J_{(1)}=dX+[J_{(0)} , X] ,\quad J_{(2)}=[J_{(1)},X],\quad J_{(3)}=[J_{(2)},X], \cdots .$$

We plug these expansions into (\ref{psaction}) and determine $S_2$ to be
\beq\label{stwo} S_2  =  \langle \half \p X_2 \pb X_2  + \p X_1 \pb X_3 + \half J_{(0)} ( [ X_1 , \pb X_1 ] + [ X_3 , \p X_3 ] ) \rangle \eeq
$$ + \langle \half J_{(0)} \left [ [ J_{(0)} ,   X_2 ] , X_2 \right]+\o\pb\l + \oh\p\lh \rangle   .$$
To verify conformal invariance, we need the action up to order $3$ because of (\ref{dSord}). Note that $\d_1 S_2$ contains quantum fluctuations of order $2$, then we have to search for terms of order $2$ in $\d_0 S_3$. Since, the $S_3$ is of order $3$, its variation is of the form $\left< X X \d X \right>$. Then, to get terms of order $2$ here, we need variations of $X$ which are independent of the quantum fluctuation $X$. The only one are $X_2$. Therefore, we determine the terms involving $X_2$ only in $S_3$. They are 
\beq\label{sthree} {1\over \sqrt{\alpha'}}S_3  =  \langle -{1\over 8}  X_1  ( [ \p X_1  , \pb X_2 ]  + [ \p X_2  , \pb X_1 ]   ) + {1\over 4}  X_2  ( [ \p X_3  , \pb X_3 ]  - [ \p X_1  , \pb X_1 ]   )  \eeq
$$
+ {1\over 8}  X_3  ( [ \p X_2  , \pb X_3 ]  + [ \p X_3  , \pb X_2 ]   )  \rangle $$
$$
+\langle J_{(0)} \left( {1\over 3} \left[ [ \p X_2 + \pb X_2 , X_2 ] , X_2 \right]
+{1\over{24}} \left[ [ 5\p X_1 + 11\pb X_1 , X_3 ] , X_2 \right] \right) \rangle $$
$$
+\langle J_{(0)} \left( {1\over{24}} \left[ [ 11\p X_3 + 5\pb X_3 , X_1 ] , X_2 \right]
+{1\over{24}} \left[ [ 5\p X_2 -\pb X_2 , X_3 ] , X_1 \right] \right) \rangle $$
$$
+\langle J_{(0)} \left( {1\over 6} \left[ [ 2\p X_3 - \pb X_3 , X_2 ] , X_1 \right]
-{1\over 6} \left[ [ \p X_1 - 2\pb X_1 , X_2 ] , X_3 \right] \right) \rangle $$
$$
-\langle J_{(0)} \left({1\over{24}} \left[ [ \p X_2 - 5 \pb X_2 , X_1 ] , X_3 \right] \right) \rangle $$
$$
+ \langle {1\over 3} [ J_{(0)} , X_1 ] \left( \left[ [ J_{(0)} , X_2 ] , X_1 \right] + \left[ [ J_{(0)} , X_1 ] , X_2 \right] \right) \rangle $$
$$
- \langle {1\over 6} [ J_{(0)} , X_2 ] \left( \left[ [ J_{(0)} , X_3 ] , X_3 \right] + \left[ [ J_{(0)} , X_1 ] , X_1 \right] \right) \rangle $$
$$
+ \langle {1\over 3} [ J_{(0)} , X_3 ] \left( \left[ [ J_{(0)} , X_2 ] , X_3 \right] + \left[ [ J_{(0)} , X_3 ] , X_2 \right] \right) \rangle $$
$$
+\langle J_{(0)} [ X_2 , N + \Nh ] \rangle .$$
Note that the first two lines in (\ref{sthree}) becomes, after integrating by parts,

$$
\langle \half  X_2  ( [ \p X_3  , \pb X_3 ]  - [ \p X_1  , \pb X_1 ]   ) \rangle .$$
Also note that
$$
\langle [ J_{(0)} , X_1 ]  \left[ [ J_{(0)} , X_1 ] , X_2 \right] \rangle =  \langle [ J_{(0)} , X_3 ]  \left[ [ J_{(0)} , X_3 ] , X_2 \right] \rangle =  0 .$$

Apart from being important to the symmetries of the action, these cubic terms are necessary to compute the canonical momenta 
for some variables. 

\section{A BPS background}

The background around which we choose to quantize the string is given by
\beq
g_0 = e^{\alpha'{\cal E}\tau T} e^{\alpha' {\cal J} \tau J}.
\eeq
It describes a point-like string rotating along an equator in $S^5$. The only non vanishing left invariant current is 
\beq J_\tau = g^{-1}_0 \p_\tau g=\alpha' {\cal E} T + \alpha' {\cal J} J.\eeq 

One can see immediately see that such classical configurations satisfy the equations of motion. 
The classical Virasoro constraint for such configuration reads
\beq\label{classvir}
T+\bar T={1\over{2\alpha'}}\textrm{Str}\, \tilde J_\tau \tilde J_\tau={\alpha'\over2}( -{\cal E}^2+ {\cal J}^2)=0.
\eeq
and it implies that ${\cal E} = |{\cal J}|$, which is the usual BPS condition. Later we will see that there are quantum corrections to this constraint.  We can also calculate the value of the conserved current
\beq j_\tau = \frac{1}{2\pi\alpha'}g J_\tau g^{-1} =\frac{1}{2\pi}( {\cal E} T + {\cal J} J), \eeq
the charges are given by 
\beq {\mathsf E} = -\oint d\sigma \Tr[ T j_\tau]=\frac{1}{2\pi}\oint d\sigma {\cal E}={\cal E},\quad  {\mathsf J} = \oint d\sigma \Tr[ \J j_\tau]=\frac{1}{2\pi}\oint d\sigma {\cal J} ={\cal J}.
\eeq
which confirms the claim that this is a BPS solution, {\em i.e.} ${\mathsf E}-{\mathsf J}=0$, when ${\cal E} = {\cal J}$. From now on we will set ${\cal E}={\cal J}$. The BMN limit is 
$$\alpha'\to 0 \; , \; {\cal J}\to\infty \; , \; \alpha'{\cal J} = {\rm finite}.$$

\subsection{Expansion of the action}

Now that we know the background, we would like to know what is the spectrum of quantum fluctuations around it. In order to do that, we use the background field quantization. The quantum coset element is 
\beq g = g_0 e^{\sqrt{\alpha'}X}, {~~~\rm where ~~~} X= t T+ \phi J+ X^A P_A+X^I P_I + \Theta^a Q_a+\Theta^{\dot a}Q_{\dot a} + \hat\Theta^a\hat Q_a+ \hat\Theta^{\dot a}\hat Q_{\dot a}.
\eeq

Using (\ref{stwo}) and the conventions in Appendix A we get the pure spinor action for the superstring on a pp-wave background
\beq\label{ppaction}
S = \frac{1}{\pi} \int d^2 z \Big\{ \half \left(- \partial t \bar \partial t + \partial \phi \bar \partial \phi + \partial X^i \bar \partial X^i -\frac{\alpha'^2{\cal J}^2}{4} X^i X^i \right) + \qquad \qquad \qquad 
\eeq
\[
\quad \qquad \qquad \qquad \qquad \qquad + \partial \hat\Theta^{\hat a} \bar \partial \Theta^{b} \Pi_{\hat a b} - i\frac{{\cal J}\alpha' }{4} \left(\Theta^a \delta_{ab} \bar\partial \Theta^b + \hat\Theta^{\hat a} \delta_{\hat a\hat b} \partial \hat\Theta^{\hat b} \right) + \omega_a  \bar\partial \lambda^a  + \bar\omega_{\hat a} \partial \bar\lambda^{\hat a} + 
\]
\[
\qquad \qquad \qquad \qquad \qquad \qquad \qquad \qquad \qquad \qquad \qquad \qquad \qquad + \partial \hat\Psi^{\dot {a}} \bar \partial \Psi^{\dot b} \Pi_{\dot {\hat a}\dot b} + \omega_{\dot a}  \bar\partial \lambda^{\dot a}  + \bar\omega_{\dot {\hat a}} \partial \bar\lambda^{\dot {\hat a}} \Big\} 
\]
where $i = (A,I)$ and we have renamed $\Theta^{\dot a}\to\Psi^{\dot a}$ and $\hat\Theta^{\dot{\hat a}}\to \hat\Psi^{\dot a}$. Note that the ghost action is just the same as in flat space. This is so because we are keeping only second order terms, we don't have any contribution from $[J_0, \lambda]$ or $N \bar N$. These are the leading terms for the action at the BMN limit. However, we are going to 
see that the cubic action (\ref{sthree}) still contributes when one computes the conjugate momenta. 

\subsection{Equations of Motion and Quantization}

The equations of motion that come from the action  (\ref{ppaction}) are
\beq \partial\bar\partial t = \partial\bar\partial \phi = \Big(4\partial\bar\partial +{\cal J}^2\alpha'^2\Big)X^I=\partial\bar\partial \Psi^{\dot a}=\partial\bar\partial\hat\Psi^{\dot a}= 0 \eeq
\beq \partial\bar\partial \Theta +i\frac{J\alpha'}{2}\Pi\partial\hat\Theta=\partial\bar\partial \hat\Theta -i\frac{{\cal J}\alpha'}{2}\Pi\bar\partial\Theta=0 \eeq

The solutions for the bosonic coordinates are the standard ones 
\beq t=  t_0 + p_t\tau +\sum_{n\neq 0} ( t_n f_n + \bar t_n \bar f_n),\quad \phi=  \phi_0 + p_\phi\tau +\sum_{n\neq 0} ( \phi_n f_n + \bar \phi_n \bar f_n) .\eeq
\beq \Psi = \psi - \Pi p_{\hat\psi} \tau + \sum_{n\neq 0} ( \psi_n f_n + \bar \psi_n \bar f_n),\quad \hat\Psi = \hat\psi+ \Pi p_{\psi} \tau + \sum_{n\neq 0} ( \hat\psi_n f_n + \bar{\hat \psi}_n \bar f_n)\eeq
\beq X^i= \cos(J\alpha'\tau)(a_+^i + a_-^i) +i\frac{1}{J\alpha'}\sin(J\alpha'\tau)(a^i_+ - a^i_-) +\sum_{n\neq 0} ( x_n^i g_n + \bar x_n^i \bar g_n) \eeq
\beq \Theta= \cos(J\alpha'\tau) \theta_0 -\sin(J\alpha'\tau) \Pi\hat\theta_0 +\sum_{n\neq 0} ( \theta_n g_n -\frac{i}{J\alpha'}(\omega_n-n)\Pi \hat\theta_n \bar g_n) + \sum_n \vartheta_n f_n\eeq
\beq \hat\Theta= \cos(J\alpha'\tau)\hat\theta_0 +\sin(J\alpha'\tau) \Pi\theta_0 +\sum_{n\neq 0} ( \hat\theta_n \bar g_n +\frac{i}{J\alpha'}(\omega_n-n)\Pi \theta_n  g_n)+\sum_n\hat\vartheta_n \bar f_n \eeq
\beq \lambda = \sum_n\lambda_n  f_n,\;\; \omega=\sum_n\omega_n  f_n,\;\; \bar\lambda=\sum_n\bar\lambda_n \bar f_n, \;\; \bar\omega=\sum_n\bar\omega_n \bar f_n.
\eeq
where
 $$f_n =e^{-in(\tau-\sigma)}\; ,\; \bar f_n =e^{-in(\tau+\sigma)}\; ,\; g_n =e^{-i(\omega_n\tau-n\sigma)}\; ,\; \bar g_n =e^{-i(\omega_n\tau+n\sigma)}\; ,\; \omega_n= \pm\sqrt{({\cal J}\alpha')^2+n^2}.$$

The conjugate momenta for the quantum fluctuations are computed using the standard definition in terms of the variation of the action with respect to their time derivatives\footnote{For fermions the definition is with the right derivative.}. For the transverse fields and spinors we have
\beq P^i = \partial_\tau X^i\; , \; P_\Psi = \partial_\tau \hat\Psi \Pi \; ,\; P_{\hat\Psi}=-\partial_\tau\Psi \Pi,\quad P_\Theta= \partial_\tau \hat \Theta \Pi -\frac{J\alpha'}{2}\Theta,\quad P_{\hat\Theta}= -\partial_\tau \Theta\Pi - \frac{J\alpha'}{2}\hat\Theta \eeq
The system is quantized using standard equal-time commutation relations
\beq [P_i,(\tau,\sigma),X^j(\tau,\sigma')]=-i \delta_i^j \delta(\sigma-\sigma'),\quad [P_t,(\tau,\sigma),t(\tau,\sigma')]=-i \delta(\sigma-\sigma'),\eeq
\beq [P_\phi,(\tau,\sigma),\phi(\tau,\sigma')]=-i \delta(\sigma-\sigma'), \quad \{P_\psi,(\tau,\sigma),\Psi(\tau,\sigma')\}=i \delta(\sigma-\sigma'),\eeq
\beq \{ P_\Theta,(\tau,\sigma),\Theta(\tau,\sigma')\}=i \delta(\sigma-\sigma'),\quad \{ P_{\hat\Theta},(\tau,\sigma),\hat\Theta(\tau,\sigma')\}=i \delta(\sigma-\sigma'). \eeq

 The only subtle point is that, as we will see shortly, the stress-energy tensor has a liner term in $\partial t$ and $\partial \phi$. Because of this, when computing their conjugate momenta, we have to go to cubic order of quantum fields (\ref{sthree}) . These higher order corrections are interpreted as corrections to the time derivatives of $t$ and $\phi$ when we replace them with the momenta. From the cubic part of the action we find 
\beq\label{momt} P_t = -\p_\tau t + \sqrt{\alpha'}(\a' {\cal J}  X_A X_A - {i\over 2} ( \Theta_a \p_\s \Theta_a + \Psi_\add \p_\s \Psi_\add - \hat\Theta_a \p_\s \hat\Theta_a - \hat\Theta_\add \p_\s \hat\Theta_\add ) +{i\over 2} \a' {\cal J} \Pi_{\add\bdd} \Theta_\add \hat\Theta_\bdd ),\eeq
\beq\label{momphi} P_\phi = \p_\tau \phi - \sqrt{\alpha'}(\a' {\cal J}  X_I X_I - {i\over 2} ( \Theta_a \p_\s \Theta_a - \Psi_\add \p_\s \Psi_\add - \hat\Theta_a \p_\s \hat\Theta_a + \hat\Theta_\add \p_\s \hat\Theta_\add ) -{i\over 2} \a' {\cal J} \Pi_{\add\bdd} \Theta_\add \hat\Theta_\bdd ).\eeq
These equations we be used to replace $\tau$ derivatives, and are interpreted as corrections to $\tau$ derivatives, not the momenta. 

The quantum conserved charges are calculated from
$$j_\tau = \frac{1}{2\pi\alpha'} g[  (J_2)_\tau  +  (J_1)_\tau +  (J_3)_\tau  + N +\bar N]g^{-1}$$
using the classical solution and the quantum fluctuations. The charge of a particular generator $T_A$  is given by 
\beq Q_A = \frac{1}{2\pi\alpha'}\oint d\sigma \Tr[ T_A g[  (J_2)_\tau  +  (J_1)_\tau +  (J_3)_\tau  + N +\bar N]g^{-1}]\eeq
The charges $ {\mathsf E}$ and $ {\mathsf J}$ are straightforward to compute up to second order in quantum fields 
\beq {\mathsf E} = {\cal J} +\frac{1}{2\pi\alpha'}\oint d\sigma \Big[\sqrt{\alpha'}\partial_\tau X^0 +\alpha' \delta_{ab}(\Theta^a\partial_\tau \Theta^b+\w\Theta^a\partial_\tau \w\Theta^b) +\alpha'\delta_{\dot a\dot b}(\Psi^{\dot a}\partial_\tau \Psi^{\dot b}+\w\Psi^{\dot a}\partial_\tau \w\Psi^{\dot b}) \Big],\eeq
$${\mathsf J} = {\cal J} +\frac{1}{2\pi\alpha'}\oint d\sigma\Big[\sqrt{\alpha'} \partial_\tau X^5 +\alpha'\delta_{ab}(\Theta^a\partial_\tau \Theta^b+\w\Theta^a\partial_\tau \w\Theta^b) -\alpha'\delta_{\dot a\dot b}(\Psi^{\dot a}\partial_\tau \Psi^{\dot b}+\w\Psi^{\dot a}\partial_\tau \w\Psi^{\dot b}) \Big]
$$
In the BMN limit only the constant and first terms contribute to these charges. So the leading contribution to $ {\mathsf E}- {\mathsf J}$ is 
\beq  {\mathsf E}- {\mathsf J}= \frac{1}{2\pi\sqrt{\alpha'}}\oint d\sigma [ P_t - P_\phi]= \frac{1}{\sqrt{\alpha'}}[p_t-p_\phi].
\eeq

\subsection{Spectrum}

In the BMN limit we have a free field theory with massless and massive excitations. To know the spectrum, we have to impose BRST and Virasoro conditions.  The stress-energy tensor up to quadratic order in quantum fields in this background is given by:
\begin{eqnarray}
T & = & \partial X^m \partial X^n \eta_{mn} - \frac{1}{\sqrt{\alpha'}}({\cal J}\alpha')\partial (t-\phi) - \frac{({\cal J}\alpha')^2}{4} X^i X^i - \\
& & - 2\partial\Psi^{\dot a}\partial\hat\Psi^{\dot { b}} \Pi_{\dot a\dot {b}} ~~ - ~~ 2\partial\Theta^{a}\partial\hat\Theta^{b} \Pi_{ab} - i\frac{{\cal J}\alpha'}{2}\Theta^a\delta_{ab}\partial\Theta^b - i\frac{{\cal J}\alpha'}{2}\hat \Theta^{a}\delta_{a b}\partial\hat \Theta^{b}\nonumber \\
\bar T & = & \bar\partial X^m \bar\partial X^n \eta_{mn} - \frac{1}{\sqrt{\alpha'}}({\cal J}\alpha')\bar\partial (t-\phi) - \frac{({\cal J}\alpha')^2}{4} X^i X^i - \\
& & - 2\bar\partial\Psi^{\dot a}\bar\partial\hat\Psi^{\dot { b}} \Pi_{\dot a\dot { b}} ~~ - ~~ 2\bar\partial\Theta^{a}\bar\partial\hat\Theta^{b} \Pi_{ab} - i\frac{{\cal J}\alpha'}{2}\Theta^a\delta_{ab}\bar\partial\Theta^b - i\frac{{\cal J}\alpha'}{2}\hat \Theta^{ a}\delta_{ a b}\bar\partial\hat \Theta^{ b}\nonumber
\end{eqnarray}

Using (\ref{momt}) and (\ref{momphi}) the replace time derivatives of $t$ and $\phi$ and the mode expansion, the zero mode of the Virasoro constraint is given by 
\beq L_0 + \bar L_0 = \oint d\sigma (T + \bar T) = 4{\cal J}\alpha' 
 -\frac{1}{2}p_t^2 +\frac{1}{2}p_\phi^2 + {\cal J}\sqrt{\alpha'}(- p_t + p_\phi) +p_\psi p_{\hat\psi}+\eeq
 $$ \theta_0\Pi\hat\theta_0 + {\cal J}\alpha'a^i_+a^i_- +  \sum_{n>0} \omega_n( x^i_{-n}x^i_n +  \bar x^i_{-n}\bar x^i_n + \theta_{-n}^a \theta_n^a + \hat\theta_{-n}^a\hat\theta_n^a) +$$
 $$ \sum_{n>0}n(t_{-n}t_n + \bar t_{-n}\bar t_n + \phi_{-n}\phi_{n} + \bar\phi_{-n}\bar\phi_{n} + \psi_{-n} \hat\psi_{n} + \hat\psi_{-n}\psi_n) 
$$
We see that the energy momentum tensor just counts the energy and the number of modes on some particular state. The normal ordering constant $4J\alpha'$ comes from the eight massive bosons. The massive spinors do not contribute. As usual, physical states have to be annihilated by both $T$ and $\bar T$.  Another requirement for physical states is that they are annihilated 
by the BRST charge, which up to quadratic order is 
\beq Q = \oint d\sigma\Tr[ \lambda J_3 +\bar\lambda\bar J_1] = \oint d\sigma\Big[ \Pi_{\dot a\dot b}\lambda^{\dot a}\partial\Theta^{\dot b} +\Pi_{\dot a\dot b}\hat\lambda^{\dot a}\bar\partial\hat\Theta^{\dot b} +
\eeq
$$ \Pi_{ab}\lambda^a( \partial \Theta^b +\frac{J\alpha'}{2}\Pi^b{}_c\hat\Theta^c) +\Pi_{ab}\hat\lambda^a( \bar\partial \hat\Theta^b -\frac{J\alpha'}{2}\Pi^b{}_c\Theta^c)\Big].$$
Inserting the mode expansion into the BRST charge we see that the massive fermionic modes decouple, and only the massless fermions remain.  This means that the all massive modes create physical states and the massless ones decouple from the spectrum.\footnote{We will see bellow that the full spectrum can be constructed using only a scalar combination of ghost vacuum.} At this order in quantum fields the bosonic fluctuations do not appear in the BRST charge. The restriction imposed on these fields will come from the Virasoro condition. 

We define the vacuum $|0\rangle$ to be annihilated by all positive modes. Excited states are created by acting with other modes. Physical states should have ghost number $(1,1)$. The simplest one was introduced by Berkovits \cite{simp} and describes the radius modulus. In the notation of the previous section this state is given by $(\lambda^a\bar\lambda^b\Pi_{ab} +\lambda^{\dot a}\bar\lambda^{\dot b}\Pi_{\dot a\dot b})|0\rangle$ and we will denote it by $|\lambda\bar\lambda\rangle$. 
Acting with the charges we have ${\mathsf E} |\lambda\bar\lambda\rangle=J |\lambda\bar\lambda\rangle$
and ${\mathsf J} |\lambda\bar\lambda\rangle=J |\lambda\bar\lambda\rangle$. 
The normal ordering constant in $T+\bar T$ means that this state is not annihilated by it. To fix this we consider another state
\beq |\Omega\rangle = e^{-2i\sqrt{\alpha'}(x^0- x^5)} |\lambda\bar\lambda\rangle,\eeq
computing its ${\mathsf E}-{\mathsf J}$ we find it is 4.  This is in fact consistent with the identification made in \cite{simp}. This string state corresponds to the SYM operator 
$$ Tr[F^2 Z^J]$$
Another was to compensate the Virasoro constraint it to add zero modes of the massive spinors. For instance, consider the state 

$$| {\bf 70}\rangle = \theta_0^a\theta_0^b\theta_0^c\theta_0^d|\lambda\bar\lambda\rangle,$$
it corresponds to a KK mode with ${\mathsf J}$ charge $J$ of the graviton and self-dual four form. Breaking the $SO(8)$ spinors to $SO(4)\times SO(4)$ spinors, one element of the above  multiplet corresponds to $\Tr [Z^{J+2}]$. Similarly, we can construct all ${\bf 128}+{\bf 128}$ lowest states expected. Since all massive modes of $\Theta$ decouple from the BRST charge, these are all physical states. The massless spinors are not BRST-closed in the scalar ghost vacuum. It is interesting to note that they are all constructed on the scalar combination of the pure spinor ghosts. This is a sharp contrast with the flat space spectrum. But it should be pointed out that the $J\alpha' \to 0$ limit do not correspond to flat space.
In summary, the lowest states are 
\bea\begin{array}{rrcl}
{\mathsf E}-{\mathsf J}=4: &  e^{-2i\sqrt{\alpha'}(x^0- x^5)} |\lambda\bar\lambda\rangle &\;&  {\bf 1}\\
{\mathsf E}-{\mathsf J}=3: &  \theta^a_0e^{-\frac{3}{2}i\sqrt{\alpha'}(x^0- x^5)} |\lambda\bar\lambda\rangle & & {\bf 8}\\
{\mathsf E}-{\mathsf J}=2: &  \theta^a_0\theta^b_0e^{-i\sqrt{\alpha'}(x^0- x^5)} |\lambda\bar\lambda\rangle &&  {\bf 28}\\
{\mathsf E}-{\mathsf J}=1: &  \theta^a_0\theta^b_0\theta^c_0e^{-\frac{1}{2}i\sqrt{\alpha'}(x^0- x^5)} |\lambda\bar\lambda\rangle &&  {\bf 56}\\
{\mathsf E}-{\mathsf J}=0: &  \theta^a_0\theta^b_0\theta^c_0\theta^d_0 |\lambda\bar\lambda\rangle &&  {\bf 70}
\end{array}
\eea
States with more $\theta$'s are complex conjugates of the ones above. Further states can be obtained acting with the zero modes of the transverse directions. 
The first type of excited state is the one that changes the values of $\mathsf E$ and $\mathsf J$ in such way that $\mathsf E -\mathsf J=0$. This states is 
\beq | \Omega_n \rangle = \theta^a_0\theta^b_0\theta^c_0\theta^d_0 e^{-in\sqrt{\alpha'}(x^0+x^5)}|\lambda\bar\lambda\rangle,\eeq 
where $n=1,2,...$ is quantized because $x^5$ is an angle variable. This state satisfy both Virasoro and BRST conditions. With the quantum conserved charges we can verify that 
\beq {\mathsf E} |\Omega_n \rangle = ({\cal J} +n)|\Omega\rangle,\quad {\mathsf J}|\Omega\rangle= ({\cal J} +n)|\Omega_n\rangle.\eeq
One of its components corresponds to 
$$Tr[Z^{J+2+n}].$$
In the state above, $n$ can be positive or negative. However, it cannot be of order $1/\alpha'$, since higher order terms in the action would be needed to study it. This is not surprising, since a state with such large negative $n$ would correspond to a state with finite ${\mathsf J}$ charge.
It is precisely this feature the differentiates the pure spinor quantization with light-cone GS. In GS, the $J$ charge is fixed to be the string length. Since the pure spinor case is conformal, it is natural to expect that the we can change $J$ by finite amounts. 

Higher stringy states are created by the non zero modes of the massive world sheet fields satisfying level match condition. For example, consider 

$$x^I_{-n}\bar x^J_{-n} \theta^a_0\theta^b_0\theta^c_0\theta^d_0 |\lambda\bar\lambda\rangle,$$
however, this state does not satisfy the Virasoro condition. The same was as above, we fix this adding an exponential 

$$x^I_{-n}\bar x^J_{-n}\theta^a_0\theta^b_0\theta^c_0\theta^d_0 e^{-i{\frac{\Delta}{2} \sqrt{\alpha'}(x^0- x^5)}}  |\lambda\bar\lambda\rangle,$$
where 
$$\Delta= \frac{2}{{\cal J}\alpha'} \sqrt{ ({\cal J}\alpha')^2+n^2} = 2 \sqrt{ 1+\frac{n^2}{({\cal J}\alpha')^2}}$$ 

Computing ${\mathsf E}-{\mathsf J}$, we find that is it indeed $\Delta$. Thus, we derive the full sting spectrum on a pp-wave background. In \cite{konishi}, the physical state considered was of this form, however, there the background value of $E$ was left as a parameter to be fixed using conformal invariance. One could repeat the analysis of that paper using the same strategy as the one used here.  

\section{Conclusion}
We have studied the application of the background field method to the $AdS_5\times S^5$ pure spinor superstring around  classical solution describing point-like strings. In the BMN limit the action is quadratic in the quantum fields and the spectrum can be computed exactly. The pure spinor description of the BMN limit given here resembles the standard Green-Schwarz description, as opposed to the description given in \cite{ppnathan}. There the BMN limit is described by a contraction of the $\mathfrak{psu}(2,2|4)$ algebra and the resulting action is non linear. Although the action given here is quadratic, the world-sheet and space-time symmetries act non-linearly 
on the fields. 

The main motivation for the present work was to understand the symmetries of these constant backgrounds in order to further 
understand the model proposed to compute the energy of the Konishi state in \cite {konishi}. A key assumption in \cite {konishi} is that 
conformal invariance is preserved at quantum level. Although a general argument for quantum conformal invariance was given in \cite {quanconsistency}, an explicit computation for that case remains to be done. Another possible application is to use this formulation to compute scattering amplitudes in the BMN background. As pointed out in \cite {ppnathan}, such amplitudes would be easier to compute using the pure spinor formulation.  

\section*{Acknowledgments}

We would like to thank Nathan Berkovits and Luca Mazzucato for discussions on related subjects. LIB wants to thank Yuri Aisaka for helpful discussions.  BCV also thanks William Linch for useful discussions. The work of BCV and OC is partially supported by FONDECYT project 1120263 The work of LIB was partially supported by Conselho Nacional de Desenvolvimento Cient\'{\i}fico e Tecnol\'{o}gico (CNPq) and Funda\c{c}\~{a}o de Apoio \`{a} Pesquisa do Estado do Rio Grande do Norte (FAPERN).

\appendix
\section{Conventions}
\label{app}
We will use an $so(4)\times so(4)$ decomposition of the  $\mathfrak{psu}(2,2|4)$ algebra. The generators will be denoted by:
\begin{equation}
{\cal H}_0=(M_{AB}, M_A, M_{IJ}, M_I),\quad {\cal H}_2=(T, P_A, J, P_I),\quad {\cal H}_1=(Q_a, Q_\add),\quad {\cal H}_3=(\Qh_a, \Qh_\add) .
\end{equation}

The non zero commutators are given by 

\begin{equation}
[M_{AB}, M_{CD}] = - \d_{A[C} M_{D]B} + \d_{A[C} M_{D]B} ,\quad [M_{AB}, M_C] = \d_{C[A} M_{B]} ,\quad 
[M_A, M_B] = -M_{AB} ,
\end{equation}
\begin{equation}
[M_{IJ}, M_{KL}] = - \d_{I[K} M_{L]J} + \d_{J[K} M_{L]I} ,\quad [M_{IJ}, M_K] = \d_{K[I} M_{J]} ,\quad [M_I, M_J] = M_{IJ} .
\end{equation}
\begin{equation}
[T,P_A] = -M_A,\quad [P_A,P_B]= -M_{AB},\quad [J,P_I]= M_I,\quad [P_I,P_J]= M_{IJ}.
\end{equation}
\begin{equation}
[M_{AB},T]= 0,\quad [M_{AB},P_C]= \d_{C[A} P_{B]},\quad [M_A,T]= -P_A ,\quad [M_A,P_B]= -\d_{AB} T,
\end{equation}
\begin{equation}
[M_{IJ},J]= 0,\quad [M_{IJ},P_K]= \d_{K[I} P_{J]} ,\quad [M_I,J]=  P_I,\quad [M_I,P_J]= -\d_{IJ} J.
\end{equation}
\begin{equation}
[M_{AB},Q_a]= \frac12 (\s_{AB})_{ab} Q_b,\quad [M_{AB},Q_\add]= \frac12 (\s_{AB})_{\add\bdd} Q_\bdd,
\end{equation}
\begin{equation}
 [M_A,Q_a]= \frac12 (\s_A)_{a\bdd} Q_\bdd,\quad [M_A,Q_\add]= \frac12 (\s_A)_{\add b} Q_b,
 \end{equation}
\begin{equation}
[M_{IJ},Q_a]= \frac12 (\s_{IJ})_{ab} Q_b,\quad [M_{IJ},Q_\add]= \frac12 (\s_{IJ})_{\add\bdd} Q_\bdd,
\end{equation}
\begin{equation}
[M_I,Q_a]= \frac12 (\s_I)_{a\bdd} Q_\bdd,\quad [M_I,Q_\add]= -\frac12 (\s_I)_{\add b} Q_b.
\end{equation}
\begin{equation}
[M_{AB},\Qh_a]= \frac12 (\s_{AB})_{ab} \Qh_b,\quad
[M_{AB},\Qh_\add]= \frac12 (\s_{AB})_{\add\bdd} \Qh_\bdd,
\end{equation}
\begin{equation}
[M_A,\Qh_a]= \frac12 (\s_A)_{a\bdd} \Qh_\bdd,\quad
[M_A,\Qh_\add]= \frac12 (\s_A)_{\add b} \Qh_b,
\end{equation}
\begin{equation}
[M_{IJ},\Qh_a]= \frac12 (\s_{IJ})_{ab} \Qh_b,\quad [M_{IJ},\Qh_\add]= \frac12 (\s_{IJ})_{\add\bdd} \Qh_\bdd,
\end{equation}
\begin{equation}
[M_I,\Qh_a]= \frac12 (\s_I)_{a\bdd} \Qh_\bdd,\quad
[M_I,\Qh_\add]= - \frac12 (\s_I)_{\add b} \Qh_b.
\end{equation}
\begin{equation}
[T,Q_a]= [ J , Q_a ] = \frac12 \Pi_{ab}\Qh_b,\quad
[T,Q_\add]= -[J,Q_\add] = \frac12 \Pi_{\add\bdd}\Qh_\bdd ,
\end{equation}
\begin{equation}
[P_A,Q_a]= \frac12 (\s_A)_{a\bdd} \Pi_{\bdd\cdd}\Qh_\cdd,\quad
[P_A,Q_\add]= \frac12 (\s_A)_{\add b} \Pi_{bc}\Qh_c,
\end{equation}
\begin{equation}
[P_I,Q_a]= \frac12 (\s_I)_{a\bdd} \Pi_{\bdd\cdd}\Qh_\cdd,\quad
[P_I,Q_\add]= \frac12 (\s_I)_{\add b} \Pi_{bc}\Qh_c.
\end{equation}
\begin{equation}
[T,\Qh_a]=[J,\Qh_a]= -\frac12 \Pi_{ab} Q_b,\quad
[T,\Qh_\add]=-[J,\Qh_\add]= -\frac12 \Pi_{\add\bdd} Q_\bdd,
\end{equation}
\begin{equation}
[P_A,\Qh_a]= -\frac12 (\s_A)_{a\bdd} \Pi_{\bdd\cdd} Q_\cdd ,\quad
[P_A,\Qh_\add]= -\frac12 (\s_A)_{\add b} \Pi_{bc} Q_c,
\end{equation}
\begin{equation}
[P_I,\Qh_a]= -\frac12 (\s_I)_{a\bdd} \Pi_{\bdd\cdd} Q_\cdd,\quad
[P_I,\Qh_\add]= -\frac12 (\s_I)_{\add b} \Pi_{bc} Q_c.
\end{equation}
\begin{equation}
\{ Q_a , Q_b \} = -i\d_{ab} ( T - J ),\quad
\{ Q_a , Q_\bdd \} = i\s^A_{a\bdd} P_A + i\s^I_{a\bdd} P_I,\quad
\{ Q_\add , Q_\bdd \} = -i\d_{\add\bdd} ( T + J ).
\end{equation}
\begin{equation}
\{ \Qh_a , \Qh_b \} = -i\d_{ab} ( T - J ),\quad
\{ \Qh_a , \Qh_\bdd \} = i\s^A_{a\bdd} P_A + i\s^I_{a\bdd} P_I,\quad
\{ \Qh_\add , \Qh_\bdd \} = -i\d_{\add\bdd} ( T + J ).
\end{equation}
\begin{equation}
\{ Q_a , \Qh_b \} = \frac{i}{2} \left( (\s^{AB})_{ac} \Pi_{cb} M_{AB} - (\s^{IJ})_{ac} \Pi_{cb} M_{IJ} \right) ,
\end{equation}
\begin{equation}
\{ Q_a , \Qh_\bdd \} = -i\left( (\s^A)_{a\cdd} \Pi_{\cdd\bdd} M_A + (\s^I)_{a\cdd} \Pi_{\cdd\bdd} M_I \right) ,
\end{equation}
\begin{equation}
\{ Q_\add , \Qh_\bdd \} = \frac{i}{2} \left( (\s^{AB})_{\add\cdd} \Pi_{\cdd\bdd} M_{AB} - (\s^{IJ})_{\add\cdd} \Pi_{\cdd\bdd} M_{IJ} \right),
\end{equation}
\begin{equation}
\{ Q_\add , \Qh_b \} = -i\left( (\s^A)_{\add c} \Pi_{cb} M_A - (\s^I)_{\add c} \Pi_{cb} M_I \right) .
\end{equation}


\begin{thebibliography}{99}

\bibitem{bena}
  I.~Bena, J.~Polchinski and R.~Roiban,
  {\it Hidden symmetries of the $AdS_5 \times S^5$ superstring,}
  Phys.\ Rev.\  D {\bf 69}, 046002 (2004)
  [arXiv:hep-th/0305116].

\bibitem{vallilo}
  B.~C.~Vallilo,
  {\it Flat currents in the classical $AdS_5 \times S^5$ pure spinor superstring,}
  JHEP {\bf 0403}, 037 (2004)
  [arXiv:hep-th/0307018].

\bibitem{benichou}
  R.~Benichou,
  {\it First-principles derivation of the AdS/CFT Y-systems,}
   [arXiv:1108.4927 [hep-th]].

\bibitem{gkv}
  N.~Gromov, V.~Kazakov, P.~Vieira,
  {\it Exact Spectrum of Anomalous Dimensions of Planar N=4 Supersymmetric Yang-Mills Theory,}
  Phys.\ Rev.\ Lett.\  {\bf 103}, 131601 (2009).
  [arXiv:0901.3753 [hep-th]].

\bibitem{Berkovits:2000fe} 
  N.~Berkovits,
  {\it Super Poincare covariant quantization of the superstring,}
  JHEP {\bf 0004}, 018 (2000)
  [hep-th/0001035].
  %%CITATION = HEP-TH/0001035;%%
  %388 citations counted in INSPIRE as of 07 Mar 2014


\bibitem{Berkovits:2000yr} 
  N.~Berkovits and O.~Chandia,
  {\it Superstring vertex operators in an $AdS_5 \times S^5$ background,}
  Nucl.\ Phys.\ B {\bf 596}, 185 (2001)
  [hep-th/0009168].


\bibitem{oneloop}
  B.~C.~Vallilo,
  {\it One loop conformal invariance of the superstring in an $AdS_5 \times S^5$ background,}
  JHEP {\bf 0212}, 042 (2002).
  [hep-th/0210064].

\bibitem{quanconsistency}
  N.~Berkovits,
  {\it Quantum consistency of the superstring in $AdS_5 \times S^5$ background,}
  JHEP {\bf 0503}, 041 (2005).
  [hep-th/0411170].

\bibitem{radius}
  L.~Mazzucato, B.~C.~Vallilo,
  {\it On the Non-renormalization of the AdS Radius,}
  JHEP {\bf 0909}, 056 (2009).
  [arXiv:0906.4572 [hep-th]].

\bibitem{algebra}
  O.~A.~Bedoya, D.~Z.~Marchioro, D.~L.~Nedel, B.~Carlini Vallilo,
  {\it Quantum Current Algebra for the $AdS_5 \times S^5$ Superstring,}
  JHEP {\bf 1008}, 026 (2010).
  [arXiv:1003.0701 [hep-th]].

\bibitem{konishi}
  B.~C.~Vallilo, L.~Mazzucato,
  {\it The Konishi multiplet at strong coupling,}
  [arXiv:1102.1219 [hep-th]].

\bibitem{radu}
  R.~Roiban, A.~A.~Tseytlin,
  {\it Semiclassical string computation of strong-coupling corrections to dimensions of operators in Konishi multiplet,}
  Nucl.\ Phys.\  {\bf B848}, 251-267 (2011).
  [arXiv:1102.1209 [hep-th]].

\bibitem{gromov}
  N.~Gromov, D.~Serban, I.~Shenderovich, D.~Volin,
  {\it Quantum folded string and integrability: From finite size effects to Konishi dimension,}
  JHEP {\bf 1108}, 046 (2011).
  [arXiv:1102.1040 [hep-th]].

\bibitem{gkv2}
  N.~Gromov, V.~Kazakov, P.~Vieira,
  {\it Exact Spectrum of Planar ${\cal N}=4$ Supersymmetric Yang-Mills Theory: Konishi Dimension at Any Coupling,}
  Phys.\ Rev.\ Lett.\  {\bf 104}, 211601 (2010).
  [arXiv:0906.4240 [hep-th]].
    
\bibitem{Aisaka:2012ud} 
  Y.~Aisaka, L.~I.~Bevilaqua and B.~C.~Vallilo,
  {\it On semiclassical analysis of pure spinor superstring in an $AdS_5$ x $S^5$ background,}
  JHEP {\bf 1209}, 068 (2012)
  [arXiv:1206.5134 [hep-th]].
  
\bibitem{Cagnazzo:2012uq} 
  A.~Cagnazzo, D.~Sorokin, A.~A.~Tseytlin and L.~Wulff,
  {\it Semiclassical equivalence of Green-Schwarz and Pure-Spinor/Hybrid formulations of superstrings in AdS(5) x S(5) and AdS(2) x S(2) x T(6),}
  J.\ Phys.\ A {\bf 46}, 065401 (2013)
  [arXiv:1211.1554 [hep-th]].
  
\bibitem{Tonin:2013uec} 
  M.~Tonin,
  {\it On Semiclassical Equivalence of Green-Schwarz and Pure Spinor Strings in AdS(5) x S(5),}
  J.\ Phys.\ A {\bf 46}, 245401 (2013)
  [arXiv:1302.2488 [hep-th]].
  
\bibitem{bmn}
  D.~E.~Berenstein, J.~M.~Maldacena and H.~S.~Nastase,
  {\it Strings in flat space and pp waves from N = 4 super Yang Mills,}
  JHEP {\bf 0204}, 013 (2002)
  [arXiv:hep-th/0202021].
  
  \bibitem{ppnathan}
  N.~Berkovits,
  {\it Conformal field theory for the superstring in a Ramond-Ramond plane wave background,}
  JHEP {\bf 0204}, 037 (2002).
  [hep-th/0203248].
  
  \bibitem{metsaev}
  R.~R.~Metsaev, A.~A.~Tseytlin,
  {\it Exactly solvable model of superstring in Ramond-Ramond plane wave background,}
  Phys.\ Rev.\  {\bf D65}, 126004 (2002).
  [hep-th/0202109].
  

 \bibitem{mikhailov1}
  A.~Mikhailov,
  {\it Symmetries of massless vertex operators in $AdS_5 \times S^5$,}
  J.\ Geom.\ Phys.\  {\bf 62}, 479 (2012)
  [arXiv:0903.5022 [hep-th]].
  %%CITATION = ARXIV:0903.5022;%%
  %9 citations counted in INSPIRE as of 07 Mar 2014


  \bibitem{mikhailov2}
  O.~A.~Bedoya, L.~I.~Bevilaqua, A.~Mikhailov, V.~O.~Rivelles,
  {\it Notes on beta-deformations of the pure spinor superstring in $AdS_5 \times S^5$,}
  Nucl.\ Phys.\  {\bf B848}, 155-215 (2011).
  [arXiv:1005.0049 [hep-th]].
  
  \bibitem{mikhailov3}
    A.~Mikhailov,
  {\it Finite dimensional vertex,}
  JHEP {\bf 1112}, 005 (2011)
  [arXiv:1105.2231 [hep-th]].
  %%CITATION = ARXIV:1105.2231;%%
  %4 citations counted in INSPIRE as of 07 Mar 2014
  
  \bibitem{Chandia:2013kja} 
  O.~Chandia, A.~Mikhailov and B.~C.~Vallilo,
  {\it A construction of integrated vertex operator in the pure spinor sigma-model in $AdS_5 \times S^5$,}
  JHEP {\bf 1311}, 124 (2013)
  [arXiv:1306.0145 [hep-th]].
  %%CITATION = ARXIV:1306.0145;%%
  %1 citations counted in INSPIRE as of 07 Mar 2014

  \bibitem{simp}
  N.~Berkovits,
  {\it Simplifying and Extending the $AdS_5 \times S^5$ Pure Spinor Formalism,}
  JHEP {\bf 0909}, 051 (2009)
  [arXiv:0812.5074 [hep-th]].
  
 % \bibitem{frolov}
 %  S.~Frolov and A.~A.~Tseytlin,
 %  {\it Semiclassical quantization of rotating superstring in $AdS_5 \times S^5$,}
 %  JHEP {\bf 0206}, 007 (2002)
 %  [arXiv:hep-th/0204226]. 


\end{thebibliography}
\end{document}